\documentclass[tinyml]{acmart}
\usepackage{graphicx}
\usepackage{url}
\graphicspath{{images/},{.}}
\DeclareGraphicsExtensions{.pdf,.jpeg,.png,.jpg}
\usepackage{multicol}
\usepackage{multirow}
\usepackage{comment}
\usepackage{xcolor}
\usepackage{paralist}
\AtBeginDocument{%
  \providecommand\BibTeX{{%
    \normalfont B\kern-0.5em{\scshape i\kern-0.25em b}\kern-0.8em\TeX}}}

\setcopyright{rightsretained}
\copyrightyear{2023}
\acmYear{2023}

\begin{document}

\title{Memory-Oriented Design-Space Exploration of Edge-AI Hardware for XR Applications}

\author{Vivek Parmar$^{1}$, Syed Shakib Sarwar$^2$, Ziyun Li$^2$, Hsien-Hsin S. Lee$^2$, Barbara De Salvo$^{2\dagger}$, and Manan Suri$^{1\dagger}$} 
\affiliation{$^1$Indian Institute of Technology Delhi, $^2$Meta Reality Labs Research \\$^\dagger$Corresponding Authors: barbarads@meta.com; manansuri@ee.iitd.ac.in}

\renewcommand{\shortauthors}{Parmar et al.}

\begin{abstract}
Low-Power Edge-AI capabilities are essential for on-device extended reality (XR) applications to support the vision of Metaverse. In this work, we investigate two representative XR workloads: (i) Hand detection and (ii) Eye segmentation, for hardware design space exploration. For both applications, we train deep neural networks and analyze the impact of quantization and hardware specific bottlenecks. Through simulations, we evaluate a CPU and two systolic inference accelerator implementations. Next, we compare these hardware solutions with advanced technology nodes. The impact of integrating state-of-the-art emerging non-volatile memory technology (STT/SOT/VGSOT MRAM) into the XR-AI inference pipeline is evaluated. {We found that significant energy benefits ($\geq$24\%) can be achieved for hand detection (IPS=10) and eye segmentation (IPS=0.1) by introducing non-volatile memory in the memory hierarchy for designs at 7nm node while meeting minimum IPS (inference per second). Moreover, we can realize substantial reduction in area ($\geq$30\%) owing to the small form factor of MRAM compared to traditional SRAM.}
\end{abstract}


\maketitle

\section{Introduction}
Extended reality (XR), \textit{i.e.}, virtual, augmented, and mixed reality is fast emerging as a key technology paradigm for the future edge and mobile systems in the incoming era of Metaverse or Omniverse. XR technology has a wide variety of applications in entertainment, communication, advertising, education, healthcare, defense, robotics, smart manufacturing, human-machine interaction, etc. XR applications are becoming more computationally intensive~\cite{illixr} which poses new challenges for designing portable XR devices and systems. The current generation portable XR devices depend extensively on high-performance compute servers to perform the heavy-lifting computation due to limitation on local device's power, compute capability, and memory capacity. This approach, however, has disadvantages such as (i) patchy and non-seamless user experiences, (ii) data transfer/network overheads, and (iii) user privacy and security concerns. Further, the explosive growth and success of techniques such as deep learning for computer vision have made computationally intensive AI-based techniques a natural use case for future XR systems~\cite{illixr}. The projected specifications of some current and future generation XR devices are shown in Table~\ref{taba}~\cite{illixr}. In certain vision-based use cases, very high-resolution ($\sim$200 MP) and high frame rates ($>$90 Hz) are required at modest power budgets ($<$1W). In this study, we perform detailed architectural design-space exploration and DTCO (design technology co-optimization) for building optimized portable XR systems while tackling some of these concerns. Our key contributions and the novel aspects are: 
(i) Two XR-specific computer vision AI workloads were analyzed: (a) Hand detection using DetNet with FPHAB\footnotemark[1] dataset and (b) Eye segmentation using UNet with OpenEDS dataset. Both models were evaluated based on full precision and post-training quantization. (ii) Benchmarking of the XR-AI applications was performed on three architectures including a general-purpose Intel-based CPU architecture and two systolic accelerator architectures: NVidia's Simba, and MIT's Eyeriss. (iii) Technology scalability study at process nodes of 28nm, 22nm, and 7nm for all three architectures was conducted and their respective EDP (energy delay product) trends were investigated. (iv) Non-volatility was introduced into the XR compute pipeline by replacing SRAM with emerging MRAM devices (STT/SOT/VGSOT MRAM) for all three architectures through two variants: (a) P0 (Weight Buffer and Global Weight Buffer replaced by MRAM), (b) P1 (all memory replaced by MRAM). (v) {Compared to SRAM-only architecture, memory power savings of 27\% with area savings of $\sim$16\% were observed for P0 variants. Correspondingly for P1 variants, memory power savings of 24\%  and area savings of $\sim$34\% were observed compared to SRAM-only variants.}
\footnote{IIT Delhi obtained and used the FPHAB dataset.}
\begin{table}[t]
  \centering
  \caption{Projected specs of state-of-the-art XR devices \cite{illixr}.}
    \begin{tabular}{|l|c|c|c|c|}
    \hline
    Metric & HTC & Ideal & Microsoft & Ideal \\ 
    & Vive Pro & VR & HoloLens2 & AR\\\hline
    Resolution (MP) & 4.6 & 200 & 4.4 & 200 \\ \hline
    Refresh rate (Hz) & 90 & 90-144 & 120 & 90-144 \\ \hline
    Motion-to-photon & \multirow{2}{*}{$<$20}   & \multirow{2}{*}{$<$20}   & \multirow{2}{*}{$<$9} & \multirow{2}{*}{$<$5} \\ 
    latency (ms) & & & & \\\hline
    Power (W) & N/A   & 1-2   & $>$7    & 0.1-0.2 \\ \hline
    \end{tabular}%
  \label{taba}%
\end{table}%

\begin{figure*}[ht]
\begin{center}
\includegraphics[width=\linewidth]{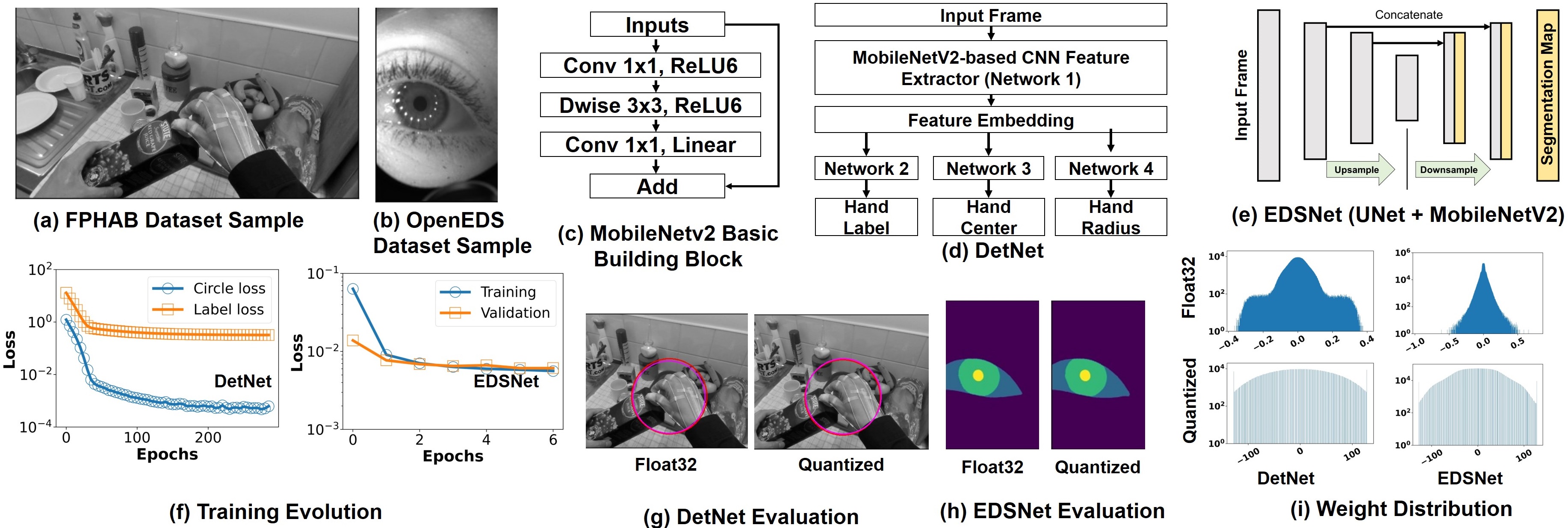}
\vspace{-0.5em}
\end{center}
   \caption{Sample images from datasets: (a) FPHAB and (b) OpenEDS. (c) MobileNetV2 building block (Inverted Residual Bottelneck~\cite{sandler2018mobilenetv2}). (d) DetNet. (e) EDSNet (UNet model + MobileNetV2 Backbone). (f) Training loss evolution. (g) DetNet evaluation, on sample image, with FP32 and INT8 precision. Red circle shows ground truth and purple shows predicted. (h) EDSNet evaluation, on sample image, with FP32 and INT8 precision. (i) Trained and quantized weight distributions for both networks.}
\label{fig_train}
\end{figure*}

\section{Analysis on Representative XR-AI Workloads}
\label{sec2}
In this section, we present algorithmic approaches for training networks used in XR-AI inference workloads of interest followed by details regarding quantization-based inference optimization. Hand detection~\cite{Garcia_Hernando_2018,Han_2020} and eye segmentation~\cite{openeds} have been heavily used as part of VR and AR headset deployment. Segmentation of ocular biometric traits of the eye region such as pupil, iris, sclera have been extensively used for studying eye movements as well as performing gaze estimation~\cite{openeds}, making it imperative to XR display designs when optimizing users' experiences. Similarly, vision-based hand tracking has been explored and adopted by commercial platforms such as Meta's Oculus Quest, HTC's VIVE, and Microsoft's Hololens as a convenient and low-friction input for XR devices to enable a seamless and less strenuous user experience. Hence, both applications are considered representative by this work for benchmarking on a variety of compute platforms for XR-AI.

\subsection{Dataset Description}
\noindent\textbf{FPHAB}. Since XR devices employ ego-centric/first-person view, we use the First Person Hand Action Benchmark (FPHAB) dataset from~\cite{Garcia_Hernando_2018} for training the hand detection network. The dataset consists 100K frames of 45 daily hand action categories, involving 26 different objects in several hand configurations. Training and validation sets consist of 52,868 and 52,212 frames, respectively. The dataset provides annotation in the form of 3D locations of 21 key joints of a hand estimated using a motion capture glove. When mapped to a 2D frame, they result in 21 keypoints. 

\noindent\textbf{OpenEDS}. For eye segmentation applications, we use the OpenEDS 2019 dataset in~\cite{openeds}. OpenEDS was collected from voluntary participants of ages between 19 and 65. The dataset contains a total of 12,759 eye images with corresponding annotation masks for key eye regions such as (a) Eyelid, (b) Iris, and (c) Pupil. The dataset is then divided into three splits, namely, (i) training (8,916), (ii) validation (2,403), and (iii) testing (1,440). The sample images from FPHAB and OpenEDS datasets are illustrated in Fig.~\ref{fig_train}(a) and Fig.~\ref{fig_train}(b).

\subsection{Network Training and Quantization}
All our neural network training experiments were performed using PyTorch. Optimized neural network architectures such as MobileNet~\cite{sandler2018mobilenetv2} have been adopted for XR applications, \textit{e.g.}, detecting hand gestures~\cite{Han_2020}. A key building block in such architectures known as inverted residual bottleneck (IRB) is shown in Fig.~\ref{fig_train}(c). The IRB helps reduce the memory footprint during inferences by not fully materializing large intermediate tensors (using depth-wise separable convolution, i.e., two layers in place of a single convolution layer), thus reducing the frequency of main memory accesses. To perform hand detection, {we trained the DetNet which is composed of a MobileNetV2-based feature extractor and three regression networks to estimate the center, the radius, and the labels of the tracked hand.} The network shown in Fig.~\ref{fig_train}(d) performs a bounding circle detection to enable the tracking of the joint movement. To train the DetNet, we first converted the keypoint annotations of FPHAB dataset to bounding circles. The center of each circle was estimated by computing the mean of x and y coordinates for each keypoint, while the radius was estimated as the maximum distance in XY plane between the center and all keypoints. The DetNet was trained over 300 epochs using AdamW optimizer. We used a combination of two loss components for the overall network training: (i) Circle loss, \textit{i.e.}, the loss in MSE {(mean square error)} for predicting center and radii of bounding circles for both hands and (ii) Label loss, \textit{i.e.}, the cross-entropy loss for predicting left hand or right hand. The training progress for each loss component is shown in Fig.~\ref{fig_train}(f). {The Circle loss is calculated as the weighted sum of the center and the radius MSE losses with a higher weight given to the center. As depicted in Fig.~\ref{fig_train}(f), the Circle loss achieves MSE values around 10$^{-3}$ within 200 epochs.}
\begin{figure*}[tb]
\begin{center}
\includegraphics[width=0.9\linewidth]{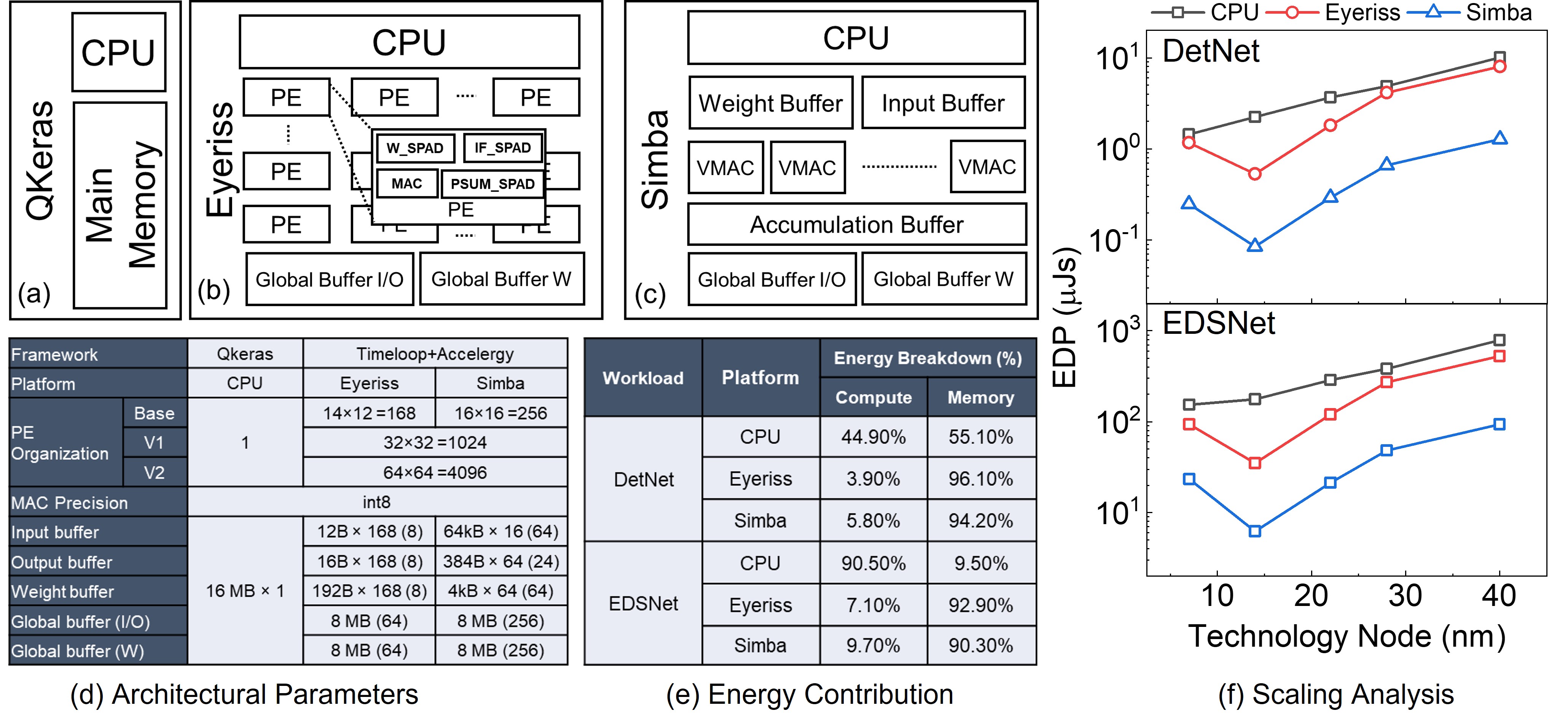}
\vspace{-0.5em}
\end{center}
   \caption{Simulated: (a) CPU + Memory in QKeras, (b) Eyeriss PE~\cite{eyeriss} and (c) Simba~\cite{simba} in Timeloop. (d) Specification of simulated architectures used in the study. Numbers mentioned in bracket indicate bus size. (e) Energy Breakdown of simulated architectures. (f) Estimated EDP for inference of DetNet and EDSNet. SRAM-only variant was estimated at 45 nm for CPU and 40 nm for Eyeriss/Simba. Scaling estimates on other nodes are based on~\cite{deepscale,tpuv4}.}
\label{fig_arch}
\end{figure*}
To perform eye segmentation, we trained EDSNet---UNet model~\cite{unet} with MobileNetV2 backbone (Fig.~\ref{fig_train}(e))---using the "segmentation models" library~\cite{Yakubovskiy:2019}. The training was performed using Adam optimizer with DiceLoss over six epochs. The training progress is shown in Fig.~\ref{fig_train}(f).  {The loss value converges within three epochs indicating high efficiency of the trained feature extractor.} Since most optimized edge AI hardware platforms can benefit from using lower precision (\textit{e.g.}, INT8), we performed post-training quantization on both models using NVIDIA's TensorRT library. The evaluation of full precision and quantized models of DetNet on samples from the dataset are visualized in Fig.~\ref{fig_train}(g). Similarly, the segmentation results on a sample image using both FP32 and quantized INT8 models of EDSNet are shown in Fig.~\ref{fig_train}(h). The weight histograms for trained and quantized models for both networks are shown in Fig.~\ref{fig_train}(i). The quantized model shows a more smooth and uniform weight distribution with discrete levels. This further helps model compression by opening possibilities for weight sharing across layers~\cite{eyeriss} The satisfactory inference results for both networks, with INT8 quantization, is exploited for hardware exploration discussed in the following sections.

\section{Implementation on Edge-AI Accelerators}
\label{sec3}

We benchmark our XR-AI workloads on three simulated architectures illustrated in Fig.~\ref{fig_arch}: {(i) a generic CPU~\cite{Coelho_2021} and two systolic inference accelerators: (ii) Eyeriss~\cite{eyeriss}, and (iii) Simba~\cite{simba}}. 
These architectural simulations help us to investigate the roles of various important design parameters such as datapath, operation mapping, parallelism, and memory hierarchy as described in Fig.~\ref{fig_arch}(d). The key difference between Eyeriss and Simba is in their memory organization. While Eyeriss heavily relies on localized memory for every PE (processing element), Simba utilizes shared buffers across rows in the form of input buffer, weight buffer, and accumulation buffer. 
For architectural workload mapping and network simulations, we used the following three frameworks: QKeras~\cite{Coelho_2021}, Timeloop~\cite{timeloop}, and Accelergy~\cite{accelergy}. In the case of QKeras (CPU), models were first translated to Keras followed by quantization using QKeras library with energy estimation based on {the operation mapping to a CPU instruction set}. QKeras maps the workload to a pure CPU architecture and provides energy estimates at 45nm node~\cite{Coelho_2021}. {QKeras also allows choices of memory configurations, they are (a) SRAM-only (b) SRAM+DRAM with writeback (c) DRAM-only. For the current study, we use SRAM-only configuration for the memory.}

Timeloop was used to estimate the cycle-wise operation mapping of the two neural network workloads on the systolic PEs based on Eyeriss (row-stationary) and Simba (weight-stationary). For using Timeloop, we exported the models from torch using the \textit{pytorch2timeloop} converter. We performed the following modifications on baselines Simba and Eyeriss to make them more relevant for the XR-AI use cases. First, DRAM was completely removed from both accelerators and SRAM global buffer size was chosen as per workload requirement shown in Fig.~\ref{fig_arch}(d). While both SRAM and DRAM are volatile memory technologies, DRAM offers a lower area/cost in contrast to that SRAM offers latency and energy benefits which are critical for such applications. Secondly, we employed Aladdin's 40nm standard cell library as a reference in place of the original 45nm one provided by Accelergy. The adoption of 40nm cell library enabled INT8 support for Eyeriss in place of the default INT16 MAC operations. Moreover, since the 40nm library offers multiple versions of modules in adders/multipliers/registers, it enables DTCO through Accelergy on the basis of energy-latency trade-offs. CACTI~\cite{fincacti} is used to estimate the energy for various SRAM buffers shown in Fig. ~\ref{fig_arch}(b) and (c). The estimated EDP for inference of both workloads---hand detection and eye segmentation---is shown in Fig.~\ref{fig_arch}(f). Apart from the baseline DRAM-free variants at 45nm/40nm, we also projected energy scaling for more advanced nodes (28nm, 22nm, and 7nm) for all three architectures. Energy and latency scaling factors used for the analysis were derived from~\cite{tpuv4,deepscale}. Scaling from the baseline technology node (45nm for CPU, 40nm for Simba/Eyeriss) leads to an energy reduction of up to 4$\times$ across all architectures. {While the systolic accelerators may have significant benefits in terms of latency, it can be observed that energy costs increase significantly as compared to a baseline CPU. In case of 7nm, Simba and Eyeriss show similar energy dissipations for EDSNet while in case of DetNet Simba shows energy savings of 11\% compared to Eyeriss. The discrepancy at 7nm observed for EDSNet can be attributed to the memory-intensive nature of the workload which benefits from row-stationary architecture of Eyeriss.}   


\begin{figure*}[t]
\begin{center}
\includegraphics[width=0.98\linewidth]{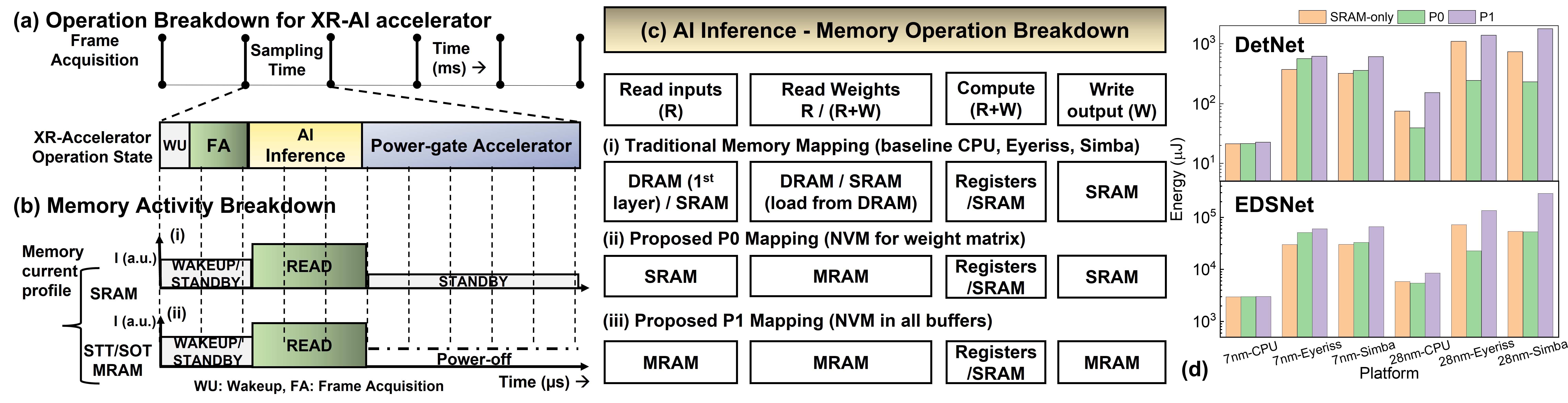}
\vspace{-0.5em}
\end{center}
   \caption{(a) Operation breakdown for XR-AI accelerator (b) Memory activity profile during XR-AI workload execution: (i) SRAM (ii) NVM. (c) Breakdown of memory specific operations in the AI Inference mode. Proposed NVM introduction strategies: (ii) PO (SRAM+MRAM) and (iii) P1 (MRAM-only). (d) Single inference energy dissipation for 9 simulated architectural variants on DetNet and EDSNet.}
\label{fig_mram}
\end{figure*}

\section{Proposed NVM-based Enhancement}
\label{sec4}
In previous sections, we explored the implication of network architecture and computing platforms in terms of {EDP. In addition to the absolute energy depicted in Fig.~\ref{fig_arch}(f), Fig.~\ref{fig_arch}(e) further analyzes the energy dissipation for the systolic architectures (Eyeriss and Simba) and indicates that memory power dissipation is far more significant than that of compute, leaving more room for optimization.} One such optimization already included was the removal of DRAM.  Furthermore, from literature it is evident that some XR-AI workloads are highly asymmetric in terms of their temporal compute requirements; {\em i.e.}, AI compute may not be executed at every cycle or uniformly with time, but rather in a sporadic manner~\cite{Han_2020}. Such peculiar compute requirements can benefit from active power-gating (\textit{e.g.,} normally-off computing) of the edge-AI accelerators to extend the battery life. An essential component required to implement power-gated/normally-off edge systems is non-volatile memory (NVM). NVM enables quick wake-up from off/sleep modes without the need of energy-hungry and time-consuming data reloads to SRAM or main memory~\cite{Suri_2019}. A major benefit of these NVMs is observed in silicon area due to use of additional BEOL process or 3D integration. As shown in~\cite{Wu_2021}, cell area reductions of up to 1.3x, 2.3x, and 2.5x can be achieved for SOT-, VGSOT-, and STT-MRAM over their high-density SRAM counterpart. Moreover, recent progress of emerging magneto-resistive/spintronic NVM (STT-MRAM, SOT-MRAM, etc.) has led to device performance comparable to that of SRAM~\cite{Wu_2021}. To assess this prospect, we performed a detailed analysis of energy dissipation of the aforementioned architectures for the two XR-AI workloads after including two state-of-the-art NVM devices, STT and SOT, in the XR-AI compute pipelines.

The temporal operation cycle of the simulated XR-AI pipeline is shown in Fig.~\ref{fig_mram}(a). It involves the execution modes in following sequence: (i) Accelerator wakeup (WU), (ii) Frame Acquisition or frame load (FA), (iii) AI Inference, and (iv) Power-Gating of Accelerator. The memory type (SRAM or NVM) used in the system will have a direct impact on the overall latency and energy. A pipeline that uses only volatile SRAM will follow the operation cycle shown in Fig.~\ref{fig_mram}(b)-(i), while an alternate pipeline that uses NVM in Fig.~\ref{fig_mram}(b)-(ii) can be powered-off during the intervals after performing inference without the need of any rewrite. Option to go in power-off mode due to non-volatility of memory leads to energy savings. We propose two strategies, P0 and P1 mappings shown in Fig.~\ref{fig_mram}(c), to adopt NVM-based pipelines in the edge devices for the XR-AI workloads. The per-inference cycle memory operation breakdown for AI inference is shown in Fig.~\ref{fig_mram}(c). In the proposed P0 mapping as shown in Fig.~\ref{fig_mram}(c)-(ii), we introduce NVM (STT and SOT) only for the weight memory. In a more aggressive variant P1 mapping, we replace all SRAM memory buffers with NVM as illustrated in Fig.~\ref{fig_mram}(c)-(iii).

\section{Results and Discussion}
\label{results}

\begin{figure}[tb]
\begin{center}
\includegraphics[width=0.98\linewidth]{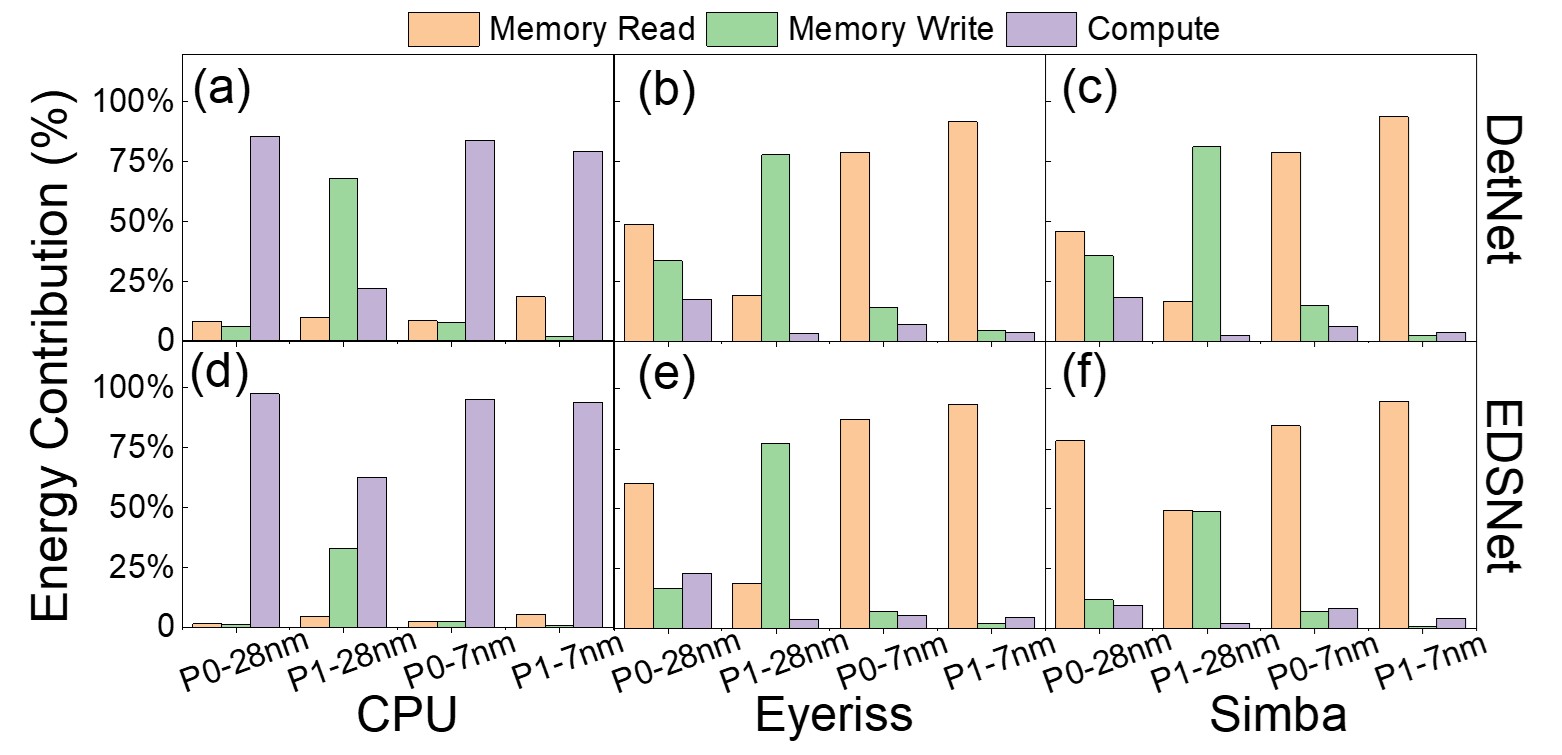}
\vspace{-0.5em}
\end{center}
\caption{Simulated energy breakdown in terms of memory and compute for NVM-based architectural variants for DetNet on: (a) CPU (b) Eyeriss (c) Simba, and  EDSNet on: (d) CPU (e) Eyeriss (f) Simba.}
\label{fig_mram2}
\end{figure}

To estimate the energy for the proposed variants P0 and P1, MRAM and SRAM macro energy characterization from recent literature is used (7nm~\cite{Wu_2021}, 28nm~\cite{Suri_2019}) along with our compute/MAC energy analysis 
The total workload energy was estimated by using operation counts based on Timeloop+Accelergy and QKeras simulations. A 64-bit memory bit-width is assumed for CPU while Timeloop employs memory bit-widths specific to the architecture (see Fig.~\ref{fig_arch}(d)). Fig.~\ref{fig_mram}(d) presents a comprehensive analysis of energy trends for both XR-AI workloads on nine different simulated architectural variants (three flavors each for CPU, Eyeriss and Simba) for two technology nodes (28nm and 7nm). For each of the three architectures, three memory flavors are considered, {\em i.e.}, SRAM only, P0: SRAM+MRAM, and P1:MRAM only. NVM technology used for 7nm estimates is VGSOT-MRAM~\cite{Wu_2021} in place of STT-MRAM. {Since the parameters used for VGSOT-MRAM are based on highly-scaled device estimates, a scaling factor based method was employed to first energy scaling in terms of SRAM. Subsequently SRAM to VGSOT-MRAM, scaling factor is employed based on literature data\cite{Wu_2021}.} Some key observations from single inference energy analysis (shown in Fig.~\ref{fig_mram}(d)) are listed below.
\begin{itemize}
    \itemsep0em 
    \item {Both P0 and P1 variants show higher energy dissipation compared to SRAM-only case at 7nm for the systolic accelerators, whereas for CPU the energy dissipation is nearly equivalent irrespective of workload.}
    \item {P1 variants show higher energy dissipation for all architectures and workloads across both nodes. This can be attributed to the asymmetric energies for read and write operation shown by MRAM as compared to SRAM.}
    \item {At 28nm, P0 variants of all architectures show energy savings compared to SRAM-only case for both workloads while a reverse trend exists at 7nm. This can be attributed to the difference in read energy costs demonstrated by STT-MRAM and VGSOT-MRAM i.e. VGSOT-MRAM is optimized for write while STT-MRAM is optimized for read.}
\end{itemize}

The detailed energy breakdown in terms of compute and memory operations (read/write) is shown in Fig.~\ref{fig_mram2}. {For all workloads and architectures based on P0 configuration and P1 at 7nm, the memory read energy dominates the memory write energy. In case of P1-28nm, this trend reverses for all architectures and workloads except for Simba with EDSNet workload. This can be attributed to the weight-stationary dataflow of Simba which results in reduced memory fetches for weights. Compute energy dominates over memory for CPU and the trend is reversed for both systolic accelerators. This can be attributed to the sequential computation dataflow employed by the CPU thus reducing unnecessary memory fetches. For P1-7nm, the memory read energy becomes overwhelmingly dominant ($\approx$ 50$\times$) in comparison to memory write energy for all architectures and workloads. This can be attributed to the fact that the VGSOT-device used for 7nm is more optimized for write as opposed to read.}

{Next, we analyze the benefits in terms of the area by introducing NVM for the systolic accelerator architectures at 7nm node. To perform area estimation, compute area was scaled as per scaling factor derived from Deepscale~\cite{deepscale}. For memory area estimates of SRAM, we utilized CACTI config files used by Accelergy with FinCACTI~\cite{fincacti} tool. Next, area scaling factors based on the feature size of a single bit-cell were derived for SRAM and VGSOT-MRAM~\cite{Wu_2021}. Using internal CACTI computations for multiple sizes of SRAM memory, periphery area factors were derived to estimate overheads at subarray, MAT, and Bank level, respectively~\cite{fincacti}. Using the above mentioned methodology, area estimates were derived for both P0 and P1 variants as summarized in Table~\ref{tabarea}. While P0 variants show marginal benefits in area ($\approx$ 16\%), P1 variants show 34\% area savings as compared to the standard SRAM-only architecture. A key reason for smaller area benefits of P0 variants can be attributed to the periphery area overhead for small memory macros. This was especially true based on the current workloads where weight memory could be optimized leading to requirements of 12 kB for storage of model weights. However for more complex workloads involving video streams, weight memory may emerge as a significant factor leading to better savings for P0.} 
\begin{table}[t]
  \centering
  \caption{{Estimation of Area Benefits on Systolic Accelerators using Proposed P0 and P1 variants at 7nm node.}}
  \label{tabarea}%
    \begin{tabular}{|c|c|c|c|c|c|}
    \hline
    \multirow{2}{*}{Architecture} & \multicolumn{3}{c|}{7 nm Area ($m m^2$)} & \multicolumn{2}{c|}{Area savings} \\
\cline{2-6}          & SRAM-only & P0    & P1    & P0 & P1 \\
    \hline
    Simba & 2.89 & 2.41 & 1.88 & 16.56\% & 34.97\% \\*
    \hline
    Eyeriss & 2.56 & 2.11 & 1.67 & 17.52\% & 34.98\% \\
    \hline
    \end{tabular}%
\end{table}%

\begin{figure}[tb]
\begin{center}
\includegraphics[width=0.98\linewidth]{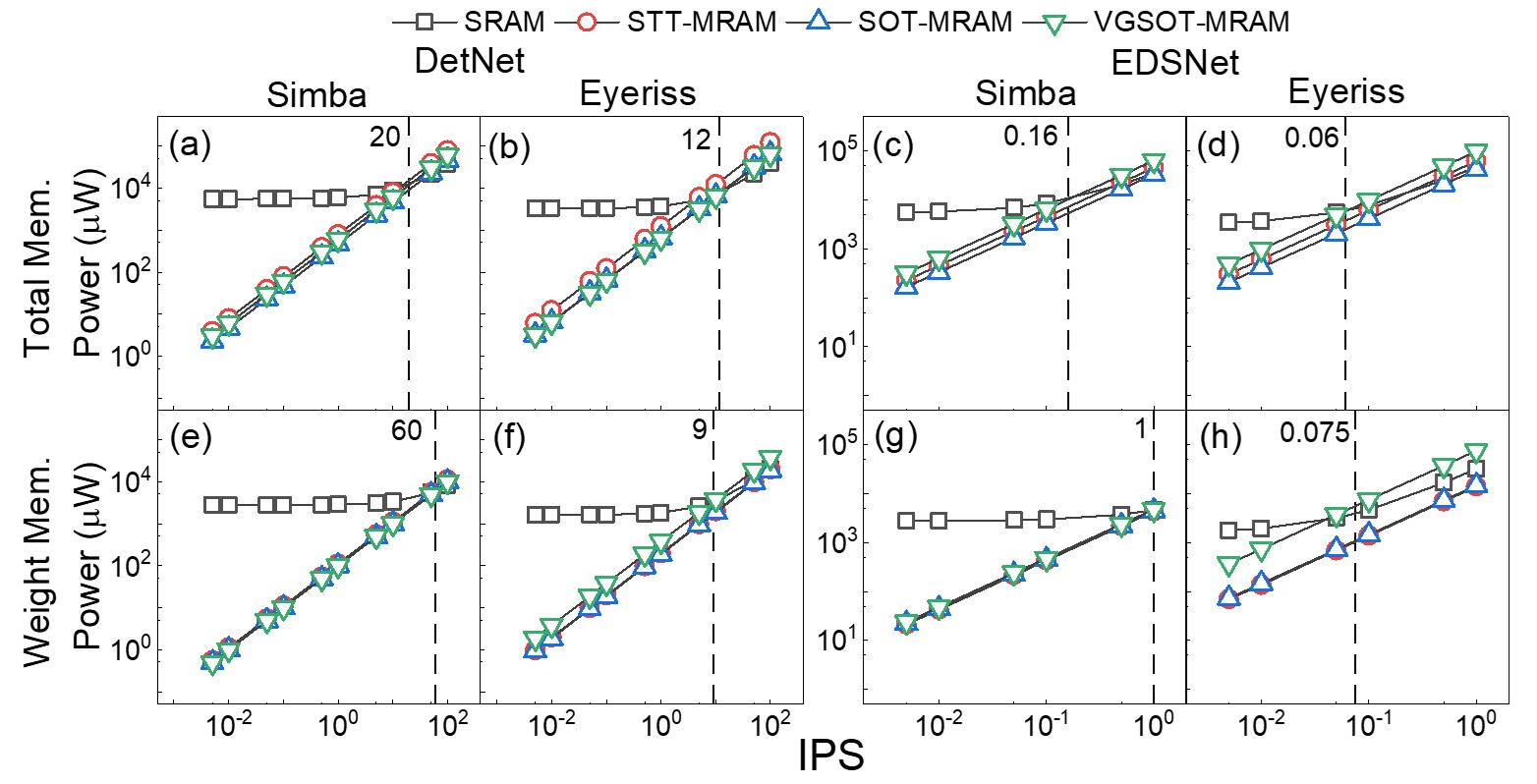}
\vspace{-0.5em}
\end{center}
   \caption{Simulated memory power vs. IPS benchmarking for proposed architectural variants utilizing SRAM, STT, SOT, and VGSOT devices. (a,b,e,f) correspond to Simba while (c,d,g,h) correspond to Eyeriss. For each case, IPS cross-over points w.r.t SRAM and VGSOT are indicated in the plots. Any IPS value below cross-over point would lead to energy-savings while using NVM. Plots (a-d) (top row) are for P1 variants while (e-h) (bottom row) are for P0 variants. {For P0 variants, cross-over points are limited based on maximum frequency supported by the memory architecture.}}
\label{fig_mram1}
\end{figure}

\begin{table}[tb]
  \centering
  \caption{IPS Analysis summary for proposed architectures using PE configuration v2 (64$\times$64).}
    \begin{tabular}{|c|c|c|c|c|c|}
    \hline
    XR-AI &  & \multicolumn{2}{c|}{Inference} & \multicolumn{2}{c|}{$P_{Mem}$ Savings} \\
    Workload & Architecture & \multicolumn{2}{c|}{Latency (ms)} & \multicolumn{2}{c|}{{@ $IPS_{min}$}} \\
\cline{3-6}  &  & P0  & P1 & P0  & P1  \\
    \hline
    DetNet & Simba  & 0.34 & 0.42 & \multicolumn{1}{c|}{27\%}  & \multicolumn{1}{c|}{31\%} \\
\cline{2-6}    {$IPS_{min}$=10} & Eyeriss   & 0.86  & 0.86 & \multicolumn{1}{c|}{-4\%}  & \multicolumn{1}{c|}{9\%} \\
    \hline
    EDSNet & Simba  & 48.57 & 60.72  & \multicolumn{1}{c|}{29\%}  & \multicolumn{1}{c|}{24\%} \\
\cline{2-6}    {$IPS_{min}$=0.1} & Eyeriss   & 45.22 & 45.22 & \multicolumn{1}{c|}{-15\%}  & \multicolumn{1}{c|}{-26\%} \\
    \hline
    \end{tabular}%
  \label{tabips}%
\end{table}%

{To analyze the impact of the asymmetric temporal compute profile of the workloads, we estimate memory power (total, weight, I/O buffer) as a function of hypothetical inference event frequency / IPS (inference per second). This metric is a direct function of the required frame rate of the application and can also assist in modeling workload for an accelerator receiving input streams from multiple sensors since the focus is on the throughput of the accelerator. In this analysis, it is assumed that accelerators can be put to sleep (power-gated) during the intervals between the completion of an inference and the arrival of the next inference request. The standby current of memory is assumed to be 100$\times$ lower compared to the read current\cite{ranica2013fdsoi} with a wakeup time of 100$\mu$s. The memory power vs IPS estimates for 7nm node using SRAM and three spintronic devices (STT, SOT, and VGSOT) are shown in Fig.~\ref{fig_mram1}. Fig.~\ref{fig_mram1}(a-d) and Fig.~\ref{fig_mram1}(e-h) show the results of memory energy savings for P1 and P0 variants, respectively. The key results on memory power savings for all combinations are summarized in Table~\ref{tabips}. In addition, the inference latency results shown in Table~\ref{tabips} reflect that Simba offers the best opportunity to exploit sleep time intervals. The latency numbers were based on the estimated cycle counts extracted from Timeloop~\cite{timeloop} multiplied by the frequency of operation for accelerator. The base frequency of compute is derived from the physically realized chips of the accelerators~\cite{eyeriss,simba} scaled down to 7nm using DeepScale~\cite{deepscale}. Operational frequency is primarily limited by memory. Hence, using peak workload-specific memory bandwidth requirements derived from Timeloop+Accelergy simulations, a relaxed operation frequency was estimated. Here we assume support for multi-cycle read and write operations using corresponding memory technology.} 

{An important point to note here is that at 7nm, all memory technologies under consideration have very low read and write latencies ($\leq$5ns) equivalent to SRAM's~\cite{Wu_2021}, thus resulting in operations running at similar inference latencies as the SRAM-only case. Here, we fix application-specific inference throughput values ($IPS_{min}$) of $\sim$10 and $\sim$0.1 for hand detection and eye segmentation applications respectively, which in the extreme case may go up to $\sim$40 and $\sim$6 respectively~\cite{KowdleRFTTDDGKK18,Feng22}}. The key observations from Memory Power vs. IPS analysis are listed below:
\begin{itemize}
    \itemsep0em 
    \item {The noticeable differences in memory power for different spintronic devices (Fig.~\ref{fig_mram1}(a-d)) can be attributed to the differences in the read and write energy for each device type (STT, SOT, VGSOT), where VGSOT has the lowest write energy but higher read energy.} 
    \item {In the case of P0 variants shown in (Fig.~\ref{fig_mram1}(e-h)), it can be observed that achievable cut-off IPS (IPS for which SRAM and MRAM variants show equal power dissipation) with VG-SOT improves for Simba whereas it decreases for Eyeriss. This can be attributed to the smaller local weight buffers used by Eyeriss requiring increased read operations in the global weight-memory.} 
    \item {P0 variants show a clear distinction in MRAM variants for the EDSNet workload which can be attributed to the increased requirement of read operations in the weight memory due to the nature of the workload.} 
    \item {While P0 variants of Simba outperform P1 variants in terms of achievable cut-off IPS (see Fig.~\ref{fig_mram1}b and Fig.~\ref{fig_mram1}f) this comes at the cost of increased power (see Table~\ref{tabips}) and area (see Table~\ref{tabarea}). Furthermore, a hybrid memory architecture would lead to higher design complexity.} 
\end{itemize}

{From the above analysis, it can be summarized that for the scaled nodes (7nm) P1 variant outperforms P0 and SRAM-only variants for DetNet workload in terms of memory power savings as well as area when operating at lower inference rates. However, this trend is reversed in case of a read-intensive workload such as EDSNet that heavily uses the input buffer and thus reduces savings from VGSOT-MRAM which is more write-optimized. P1 variants also incur the cost of slightly higher inference latency ($\approx$20\%).  However, this can be considered inconsequential with regards to the application since the latency of the P1 variants can very well satisfy the minimum IPS requirement of the application for real-world use cases. Using the accelerators with uniquely different dataflows, we can observe that while row-stationary may be beneficial for energy savings in a conventional CMOS architecture, weight-stationary dataflow leads reduced stress on memory bandwidth. This in turn facilitates the applicability of NVM in the memory hierarchy.} This makes a case for switching to higher proportion of on-chip NVM with aggressive device scaling. {However, based on the nature of workload and IPS requirement of the application, a complete replacement of on-board volatile memory with NVM may not be the optimal choice as NVM write latency might limit the computation speed. Furthermore, given the asymmetric energy dissipation trends of read and write operations for state-of-the-art NVM devices the power benefits maybe limit. Hence, based on the exact nature of the workload (\textit{i.e.} memory read-dominated or memory write-dominated), one needs to carefully fine-tune the proportion of the splits between NVM and SRAM to achieve the optimal results.}

\section{Conclusion}
\label{conclude}
We present a detailed study on two XR-AI workloads (hand detection and eye segmentation). We first present results for network training and quantization. To perform more extensive design exploration, simulations were performed for CPU and systolic accelerators using QKeras and Timeloop+Accelergy frameworks with node-scaling analysis. Finally, we propose memory-oriented DTCO based on the use of different types of the emerging MRAM devices. We also analyze the energy benefits of introducing non-volatility in the XR compute pipeline with respect to the inference activity rates at 7nm node. 
{When MRAM NVM was introduced in the memory hierarchy,  
 memory energy savings $\geq$24\% were observed for hand detection (at IPS = 10) and eye segmentation (at IPS=0.1), respectively. Additionally, MRAM replacing SRAM leads to substantial area reduction ($\geq$30\%) due to the high density feature of MRAM technology.} 

\bibliographystyle{ACM-Reference-Format}
\bibliography{ref}
\end{document}